\documentclass[referee]{raa}            

\usepackage{graphicx,times}             
\usepackage{natbib}
\usepackage{amssymb,amsmath}
\bibpunct{(}{)}{;}{a}{}{,}

\begin{document}

   \title{Triple Range Imager and POLarimeter (TRIPOL) --- A Compact and Economical Optical 
        Imaging Polarimeter for Small Telescopes
}

   \volnopage{Vol.0 (20xx) No.0, 000--000}      
   \setcounter{page}{1}          

   \author{S. Sato   \inst{1}
   \and P.~C. Huang  \inst{2}
   \and W.~P. Chen   \inst{2}
   \and T. Zenno     \inst{1}
   \and C. Eswaraiah \inst{2}\footnote{Now at National Astronomical Observatory of China, Beijing, China}, 
   \and B.~H. Su \inst{2} 
   \and S. Abe \inst{2}\footnote{Now at Department of Aerospace Engineering, Nihon University, Japan}, 
   \and D. Kinoshita  \inst{2}
   \and J.~W. Wang    \inst{3}
   }

   \institute{  Astrophysics Department, Nagoya University, Nagoya, Japan 464-8602
       \and Graduate Institute of Astronomy, National Central University, Taoyuan Taiwan 32001
       \and National Tsin Hua University, Hsin Chu
       \vs\no
   {\small Received~~20xx month day; accepted~~20xx~~month day}}

\abstract{ 
We report the design concept and performance of a compact, light-weight, and economic imaging polarimeter, 
TRIPOL (the Triple Range Imager and POLarimeter), capable of simultaneous optical imagery and polarimetry.  
TRIPOL splits the beam from wavelength 400 to 830~nm into $g^{\prime}$-, 
$r^{\prime}$-, and $i^{\prime}$-bands with two dichroic mirrors, and measures polarization with 
an achromatic half-waveplate and a wire-grid. The simultaneity makes TRIPOL a useful tool for small 
telescopes for photometry
and polarimetry of time variable and wavelength dependent phenomena. TRIPOL is devised for a 
Cassegrain telescope of an aperture of $\sim$1~m.  This paper presents the engineering considerations 
of TRIPOL and compares the expected with the observed performance.  
Using the Lulin 1-m telescope and 100~seconds integration, the limiting magnitudes are 
$g^{\prime}\sim19.0$~mag, $r^{\prime}\sim18.5$~mag and $i^{\prime}\sim18.0$~mag with a signal-to-noise of 10, 
in agreement with design expectation.  The instrumental polarization is measured to be $\sim0.3$\% 
at three bands.  Two applications, one to the star-forming cloud IC\,5146, and the other to the young 
variable GM\,Cep, are presented as demonstration.
}
   \authorrunning{W. P. Chen et al. }            
   \titlerunning{Triple Range Imager and POLarimeter
   }  

   \maketitle

%
%
\section{Introduction}           
\label{sec:int}



Polarization provides information about a celestial object additional to that by photometry and spectroscopy 
\citep{tin96,cla10}.
Yet a polarimeter has been considered as a special instrument for an optical telescope with a small aperture size. 
Nowadays, with commercial CCD cameras and other optical and electronic components readily available with good 
performance, it becomes feasible to design and fabricate a compact, and economical polarimetric imager to be used 
for scientific programs with small telescopes.  We report on an imaging system, 
TRIPOL (Triple Range Imaging POLarimeter), capable of simultaneous imaging photometry and polarimetry 
at three optical ($g^{\prime}, r^{\prime}, i^{\prime}$) bands. 
TRIPOL was designed for a telescope with the primary mirror around one meter and located at a moderate 
observing site, under typical seeing 1 to 2 arcseconds. The telescope is assumed to have a Cassegrain 
f-ratio from F/6 to F/15, and the CCD pixel scale is from 10 to 20 $\mu$m to sample 
properly the point spread function.  The optics uses no lenses to magnify or reduce the image, and the 
elements, such as the dichroic mirrors, spectral filters, a half wave-plate, and wire-grid, are all flat 
and thin for easy optical alignment.  
TRIPOL is compact, measuring $300\times350\times250$~mm in width, length, 
and height, weighing only 15~kg including the data acquisition system, and is easy to operate. 
It was devised to an accuracy $\sim3\arcsec$ for alignment, and $\sim0.05$~mm 
for machining and positioning.  

This paper describes the performance of the first (TRIPOL1) and second (TRIPOL2) units of TRIPOL 
adapted to the Lulin One-meter Telescope (LOT) in Taiwan.  In the F/8 beam of the LOT, hence a $\sim3\degr$ 
cone-angle, effects such as spherical aberration, chromatic aberration, and astigmatism are small 
compared to the 20~$\mu$m pixel size.  We compare the design parameters with observational results on 
the polarization measurements of polarization standard stars, and demonstrate the use of TRIPOL in the star-forming 
cloud IC\,5146 and the young star GM\,Cep.

\section{Design of TRIPOL}  

\begin{figure}[htb]
	\includegraphics[width=\textwidth,angle=0]{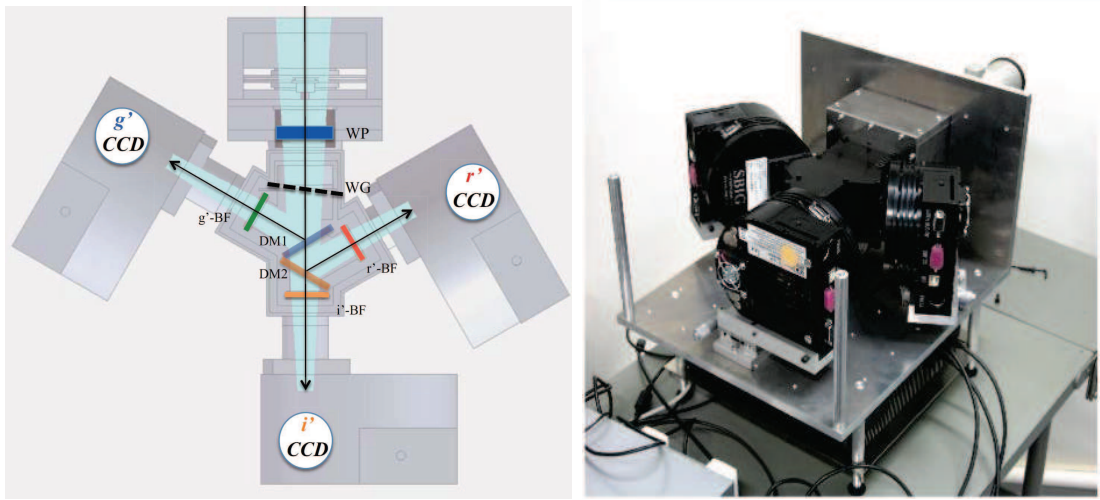}
	\caption{(Left) Layout of the optical components and CCDs of TRIPOL\,2.  Arrows show the light 
		paths. See the text for abbreviations. 
		(Right) Overview of the components with the controller computer for data acquisition 
		beneath the bottom plate. 
	}
  \label{fig:hardware}
\end{figure}

TRIPOL is composed of three parts, the polarization unit, the color-decomposition unit, and 
the data-acquisition unit, plus three CCD cameras and a desktop computer.  
The overview and layout of the optical components are shown in Figure~\ref{fig:hardware}.  
Light from the telescope passes through a half-wave plate (HWP) and a wire-grid polarizer(WG), and 
then is decomposed by two dichroic mirrors (DM1 and DM2) and band-pass filters (BPFs) into three 
($g^\prime$, $r^\prime$, $i^\prime$) channels.  The incident photons are detected and converted 
to electrons in the CCD camera with the built-in readout electronics.  

The polarization unit, consisting of a rotatable HWP and a fixed WG, working as a phase-retarder 
and polarization analyzer, respectively, is located in front of the color-decomposition unit. 
The HWP, with a size 33~mm square and thickness 3~mm, made of $SiO_{2}MgF_{2}$, is 
	procured from the optical shop, Kogaku-Giken Co. 
	We employ a commercial (from Edmund Optics Co.) WG plate of Al-wire grid, with a size 
	50~mm square and thickness 1.5~mm, sandwiched 
by thin glass plates, affording a field-of-view as wide as the detector size.  
While using birefringent materials such as a Wollaston prism would allow for, alternatively, 
a dual beam design, thus minimizing instrumental and sky effects on polarization measurements, 
our design is much more compact and economical. The WG is slightly tilted to avoid ghost images 
due to reflection glaring. 

\begin{figure}[htb]
	\centering
	\includegraphics[width=0.7\textwidth,angle=0]{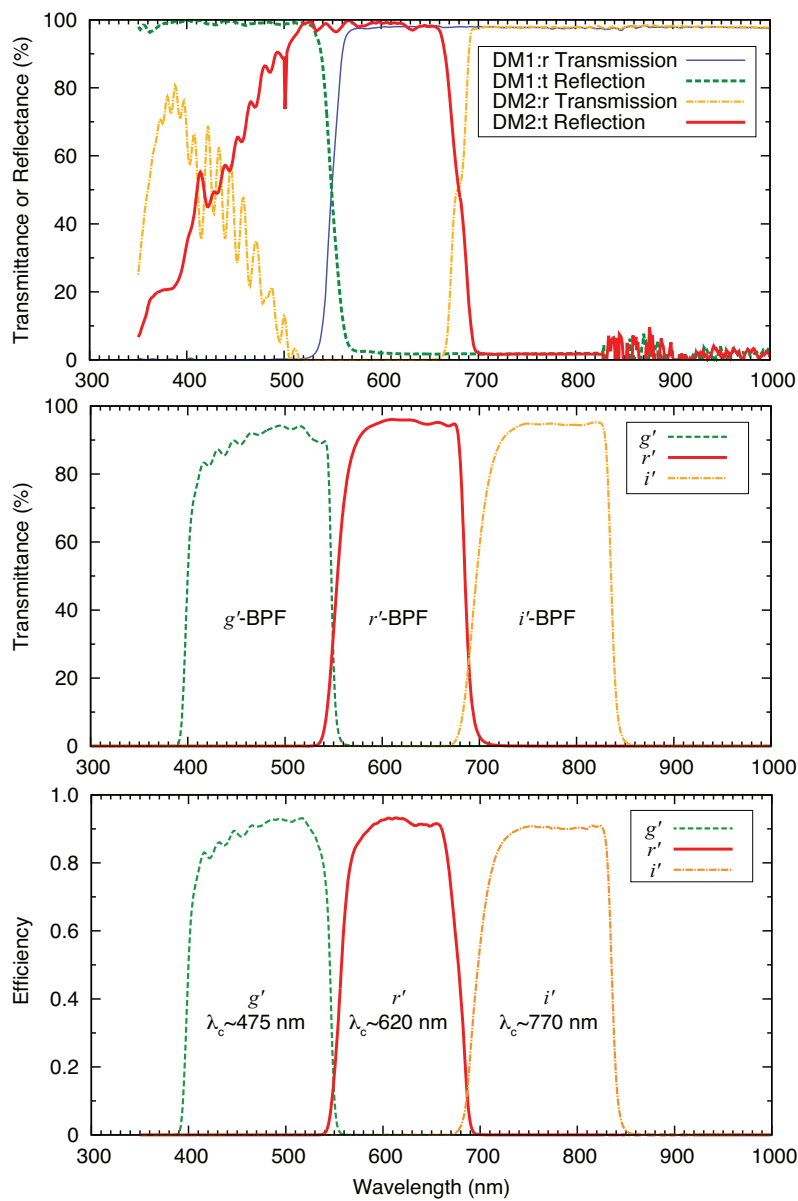}
	\caption{Transmittance of optical components for the three passbands: 
	(Top panel) Transmission/reflectance of the dichroic mirrors; 
	(Middle panel) Transmission of the bandpass filters, and
        (Bottom panel) The throughput. }
  \label{fig:trans}
\end{figure}

For the color-decomposition unit, the central wavelengths ($\lambda_0$) and bandwidths 
($\Delta\lambda$) are defined by multiplying the transmission or reflection curves of the 
DMs and BPFs for each of the $g^\prime$-, $r^\prime$-, and $i^\prime$-bands.   The spectral 
response functions of the DMs and BPFs are shown in Figure~\ref{fig:trans}.  

Even though the TRIPOL optics makes no use of mirrors or lenses with power, astigmatism 
from tilted DMs and spherical aberration from flat-parallel BPFs may still exist. 
The $g^\prime$-band optical train contains the BPF-$g^\prime$ and a CCD camera, 
the $r^\prime$-band optical train contains DM1 tilted at an angle of $30\degr$, and the 
$r^\prime$-$i^\prime$ optical train contains DM1 and DM2 at angles of $\pm30\degr$. 

Ray-tracing was executed using the software ZEMAX for classical Cassegrain-type telescopes with 
apertures $D$=0.7, 1.0, and 1.5~m, and $f$-ratios F/6, F/8, F/10, F/12.5, and F/15. 
We evaluated the tolerance of aberrations by comparing the root-mean-square (RMS) radius in 
the spot diagram with the detector pixel size and the seeing size.  It was confirmed that 
the RMS radius of the spot, due mostly to astigmatism, was smaller than 
50~$\mu$m, or $\sim$2.5 times of the pixel size, even near the corners of the detector, and 
much smaller than the seeing size, $1\farcs5$.  With various parameters for apertures and 
focal-ratios, 
the optics is found tolerable for an F/7 or a slower beam.  Astigmatism can be remedied by 
wedging DM1 and DM2 by $0\fdg18$ and $0\fdg24$ , respectively, even for a system as fast as F/7.
 
TRIPOL was designed to use commercially available CCD cameras, with the specific model 
in accordance with scientific and budgetary requirements.  TRIPOL1 and TRIPOL2 employed 
SBIG ST-9 XEi cameras using KAF-0261E plus TC-237, having $512 \times 512$ pixels, each 
with 20~$\mu$m on a side. 
The detector response shows linearity up to $\sim$50,000 counts, or about 1/3 of the full well. 
The dark current is 10~[e/s] at temperature $\sim0\degr$C, and the readout noise is 15~[e] per sampling. 

The CCD cameras are located on the bottom plate so as to align each of the array centers to the 
focal point of the telescope within an accuracy of less than 0.1~mm ($\sim$5 pixels) relative to each other. 
The SBIG ST-9 camera model satisfied our initial need for point-source targets, but the model is no longer 
available.  Subsequent TRIPOL units were upgraded to the camera model STT-8300.  
A computer (Intel DN2800MT) controls simultaneous readout of the three CCD cameras and the 
polarization units according to position angles of HWP via three USB2.0 cables. 
The overall cost of TRIPOL, excluding the cameras and the computer, was about US\$17,000\ in 2010. 
  
\section{Evaluation of performance}

In this section we evaluate the performance of TRIPOL in photometric and in polarimetric measurements.  
In each case, the engineering design parameters are compared with those measured in actual observations. 
 
\subsection{Limiting magnitudes for photometry}  

The limiting magnitudes of TRIPOL2 were measured in December 2012 using the LOT, for which 
each SBIG ST9-XEi 20~$\mu$m pixel corresponds to $0\farcs5$, 
giving a field of view of about $4\farcm7$ across.

We observed the Landolt Field 101-404 \citep{lan92} for 100~s, and analyzed the images of the 12 stars 
with a photometric aperture of $4\farcs0$, or 8 pixel in diameter, and 
derived the limiting magnitudes of 19, 18.5 and 18, for S/N$\sim10$, respectively, at 
the $g^\prime$-, $r^\prime$-, and $i^\prime$-bands.  In every band, the measured and expected values 
are in agreement with each other within the uncertainties of $\sim0.5$~mag.  
For a photometry-only observing run, the WG could be removed to gain an increase of about 
60\% incident flux.

\subsection{Efficiency and reliability of polarization measurements}

The combination of a rotatable HWP and a fixed WG, as described in section 2, follows the same design as
the near-infrared(J,H,Ks) polarimeter, SIRPOL, on the IRSF \citep[InfraRed Survey Facility,]{kan06}{}.
We describe below the performance papameters measured in laboratory, in comparison with observations of standard stars.

\subsubsection{Efficiency of the polarization devices}

\begin{figure}[htb]
  \centering
  \includegraphics[height=0.8\textheight,angle=0]{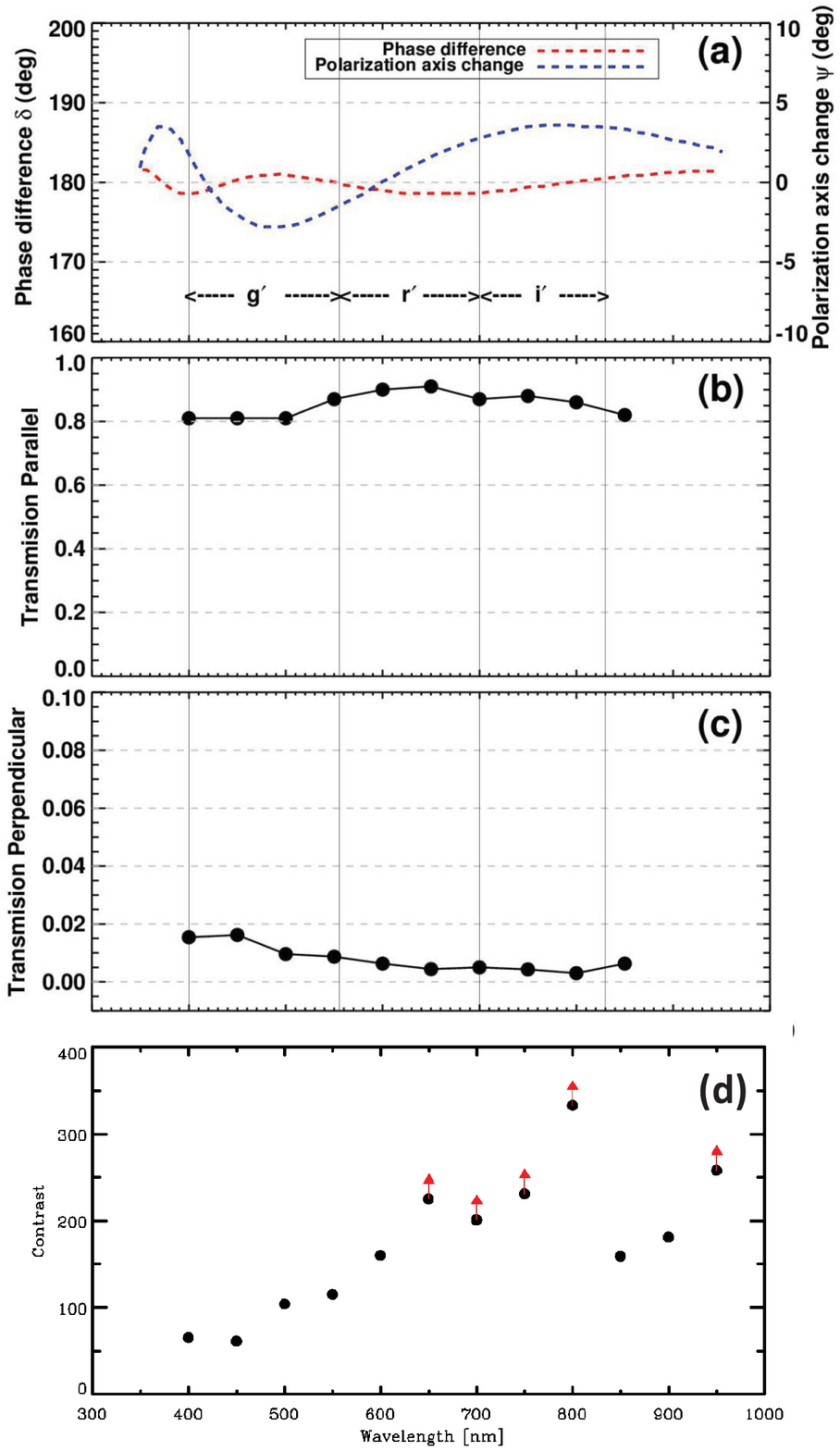}
  \caption{Performance of the TRIPOL wire-grid polarizer: (a)~The phase retardant (in unit of degree); 
	(b)~Transmittance with two polarizers in a parallel configuration; (c)~Transmittance with two polarizers 
	in a perpendicular configuration; (d)~The Contrast parameter ($T_{\rm max}/T_{\rm min}$).  
	Those with a value greater than 200, each marked with an upward arrow, are uncertain 
	because of a small number in the denominator.
	         }
  \label{fig:retar}
\end{figure}

The phase retardant of the HWP was designed and measured by Kogaku Giken Co.\ to be $180\pm2\degr$ over the 
wavelength range 400 to 950~nm (see Figure~\ref{fig:retar}(a)).  The transmittance of the WG was measured in this wavelength 
range in steps of $\Delta\lambda=50$~nm. 
Two identical WG polarizers were arranged such that one was fixed while the other was rotatable. 
When rotating relative to each other, a silicone photodiode was illuminated with a white light 
through the intermediate bandpass filters of $\Delta\lambda=50$~nm. A single rotation gives a 
double sinusoidal curve.  Fitting by a sinusoidal curve, we obtain $I(\theta)= A\sin~2(\theta-\phi)+B$, 
where $A$ is the amplitude, $B$ the residual, and $\phi$ the phase-difference. For this we parameterized  
the transmittances, $T_{\rm max} $ and $T_{\rm min}$ in parallel and perpendicular with 
each other, respectively, shown as \ref{fig:retar}(b) and \ref{fig:retar}(c). 
The contrast parameter, defined as the extinction ratio, $T_{\rm max}/T_{\rm min}$, should be as 
high as possible (infinite for a perfect polarizer), but in practice is considered satisfactory for a value 
above $\sim100$ to suppress substantially the perpendicular component of polarization, i.e., the crosstalk.  
The contrast parameter measured for TRIPOL, presented as Figure~\ref{fig:retar}(d), increases 
toward long wavelengths, and remains sufficiently high above 500~nm.


\subsubsection{Observations of Polarization Standard Stars}

The TRIPOL images were reduced by standard procedures for bias and dark 
subtraction, and correction with flatfielding.  For each polarization measurement, target frames 
acquired with each filter at four HWP positions are aligned using DAOPHOT 
(find, daomaster, and daogrow) and IRAF (geomap and geotran) packages.  Then multiple frames 
for each HWP are average-combined using IRAF/imcombine. These four images, taken at each of the four 
HWP positions, become the science images used for photometry and polarimetry.

Aperture photometry is performed using DAOFIND (for source detection with a threshold of $5\sigma$ 
of the sky variation) and PHOT (for aperture photometry) tasks of DAOPHOT for point sources.  
Typical image FWHMs for these runs varied between 2 and 4 pixels ($1\arcsec$--$2\arcsec$). 
Fluxes of each star at four positions of HWP were estimated using IRAF/DAOPHOT with an aperture 
size of 2.5 times of FWHM, as flux (and thus polarization measurements) versus aperture size 
reveals constant measurements after 2.5 times FWHM pixels. The inner and outer sky annuli are 
chosen to be 5 and 10 pixels more than the star aperture.  Fluxes at four angles are used to compute 
the Stokes parameters as follows, 
  \begin{eqnarray*} 
	I & = & 1/2 (I_{0} + I_{22.5} + I_{45} + I_{67.5}) \\
	Q & = & I_{0} - I_{45} \\  
	U & = & I_{22.5} - I_{67.5}, 
  \end{eqnarray*}
where $I_0$, $I_{22.5}$, $I_{45}$, $I_{67.5}$ are intensities at the four HWP angles in degree, 
with the corresponding error as the square-root of the sum of the square of each intensity error, i.e., 
$\delta I=\sqrt{ (\delta I_{0})^2+ (\delta I_{22.5})^2 + (\delta I_{45})^2+(\delta I_{67.5})^2 }$.  
The errors $\delta Q$, and $\delta U$ are computed similarly.

The level of polarization $P$ (in percentage) and the polarization position angle $\theta$ (in 
degree) are then derived accordingly,
\begin{eqnarray*}
       P & = & \sqrt{ Q^2 + U^2 }/I   \\
  \theta & = & 0.5 \arctan (U/Q), 
\end{eqnarray*}
for which $\delta P$, and $\delta \theta$ are estimated from the respective 
	$\delta Q$ and $\delta U$.

Because $P$ is positively defined, the derived polarization is over-estimated, 
especially for low S/N sources. To correct for this bias, the debiased value 
$P_{\rm db} = \sqrt{ P^2-(\delta P)^2 }$ \citep{war74} is computed.

\begin{figure}[htb]
  \centering
  \includegraphics[width=\textwidth]{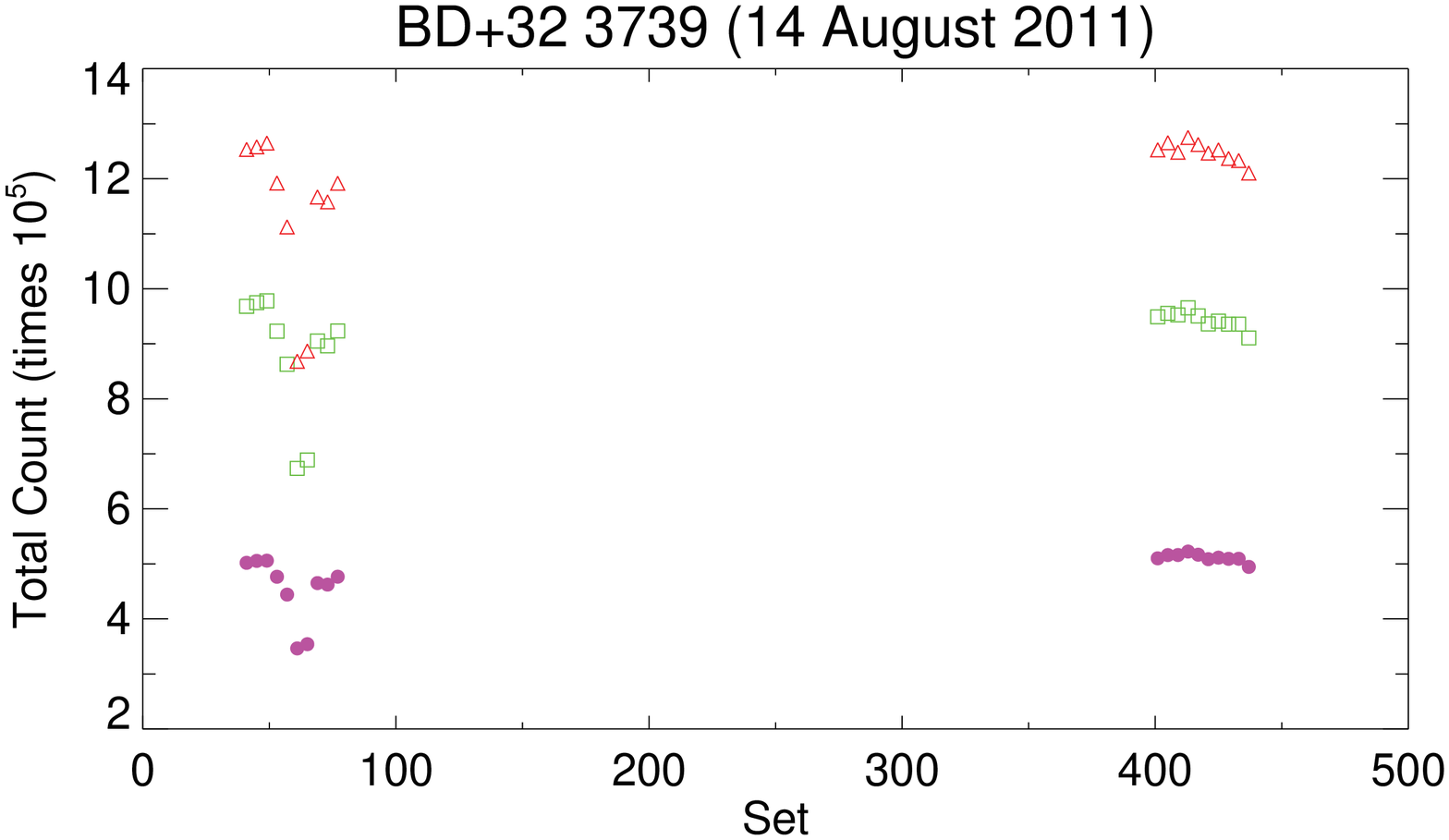}
  \caption{The total counts of BD$+$32\degr~3739 showed inferior sky 
	conditions in all $g^\prime$- (in green), $r^\prime$- (in red), and 
	$i^\prime$-bands for the first half night, whereas the sky was 
	relatively stable in the second half.
         }
  \label{fig:bd}
\end{figure}

A polarization measurement relies on photometry at different polarization angles, therefore 
all conditions pertaining to reliable photometric measurements apply.  Even under a perfect photometric 
sky, though, our observations, via a fixed sequence of images taken at 0-45-22.5-67.5 degrees, is subject to 
a small but noticeable flux drift due to airmass changes, leading to spurious polarization signals.  
Figure~\ref{fig:bd} illustrates the results observed by LOT/TRIPOL on 14 August 2011 for BD$+$32\degr~3739, 
a standard star known to have null polarization \citep{sch92}.  The total $g^{\prime}$ count, that is, the sum 
$g^{\prime}_{0} + g^{\prime}_{22.5} + g^{\prime}_{45} + g^{\prime}_{67.5}$ 
indicates varying sky conditions during the first session (a total of 10 sets of data, each set 
consisting of images at four polarization angles per filter), starting at UT~12:53 
(local time 20:53), but relatively stable skies during the second session (also with 10 sets), 
starting local time 02:12.  The 
ratio of the standard deviation of the total counts to the average counts, used as a measure of the 
sky stability, changed from about 13\% in each of the $g^{\prime}$-, $r^{\prime}$-, and $i^{\prime}$-bands 
in the first session, to about 1\% in the second session.  The data taken in the first session hence
would be discarded.

A further correction is the polarization introduced by the instrument, which is estimated by 
observing unpolarized standard stars.  The mean and standard deviation of the measured 
polarization of unpolarized standards were found to be $P_{g^\prime}= 0.27\pm0.12$\%, 
$P_{r^\prime}= 0.32\pm0.23$\%, and $P_{i^\prime}= 0.25\pm0.13$\%. 
These values, summarized in Table~\ref{tab:unpol} are considered as the instrumental polarization. 
For the unpolarized standard stars, with brightness up to ${\rm V}\sim12$~mag,  
the overall accuracy of polarization measurements with TRIPOL, is estimated to be 
$\sim0.3$\% with an uncertainty of $3\degr$ for the polarization angle.  

In every TRIPOL run, polarized standard stars are to be observed to calibrate the measured 
polarization angle to the equatorial coordinate system.  The TRIPOL measurements of known 
unpolarized standard stars and polarized stars are listed in Table~\ref{tab:unpol} and in 
Table~\ref{tab:pol}, demonstrating a general agreement with the published values, 
given the observing wavelengths are slightly different.  
An observing run was carried out exclusively for standard star calibration in 2018 October to 
assess the intranight and internight consistency of the TRIPOL measurements.  For unpolarized
standard stars, accuracy is kept to two decimal digits, and no polarization angle is listed.  
For polarized standard stars, the fractional polarization is kept to one decimal digit, with 
the polarization angle in inger to reflect the uncertainties.  In the 2018 October run, each 
target was measured a few times, and the entries in Table~\ref{tab:unpol} and Table~\ref{tab:pol} 
for each date are the average values of individual measurements and the associated errors. 
Because we relied on the standard stars to correct for the polarization angles (one offset 
per night for each angle at each band), so the values of angles scatter around the offset.    
From the observations of polarization standards, we conclude that the WG polarizer has a 
high efficiency to measure polarized light, and there is no need to correct for instrumental 
polarization except an angular offset. 

Note that only standard stars from \citet{sch92} known not to vary were selected.  In the passing
of our experiment, we found that one target, HD\,19820, however, exhibits noticeable variability 
in the polarization level, but with a relatively steady polarization angle in our measurements. 
The mechanism of the variability is unclear, but this O-type star is reported to be a binary system 
with a period of 3.366324~days \citep{hil75,hil94}. Its polarization variability requires 
further study, but in any case, using it as a standard is not advised.


\section{Scientific Demonstration}

Data acquired by TRIPOL provide simultaneous information such as flux, linear polarization and the source 
coordinates in three bands, enabling the study of the spectral energy distribution (SED), 
the color-magnitude and color-color diagrams, and polarization.  The combination of wavelength-dependence 
of polarization with the SED could distinguish various emission and propagation processes, such 
synchrotron emission, scattering or extinction.   

For imaging photometry, the time resolution of TRIPOL is as fast as about 1~s, whereas for polarimetry 
it is $\sim15$~s.  
As a single-beam instrument, TRIPOL is susceptible to polarization caused by the instrument itself, 
and to sky variations.  The effects of internal polarization are assessed by observing standard 
stars.  To mitigate the sky effects, multiple sets of observations are taken, and those with comparable 
total counts in four polarization angles are used in polarization analysis.  This compromises the 
time resolution to a few minutes, but because of the simultaneity in three bands, TRIPOL still proves 
efficient.  
TRIPOL should be especially useful to investigate variability phenomena on timescales from a few seconds 
to years or longer.  These include, but not limited to, gravitational-wave counterparts \citep{mor16}, 
gamma-ray bursts, cataclysmic variables, eclipsing binaries, Cepheids, novae, 
supernovae, blazars, Miras, and T Tauri stars \citep{che15,hua19}. 

\begin{figure}[htb]
  \centering
  \includegraphics[height=0.8\textheight]{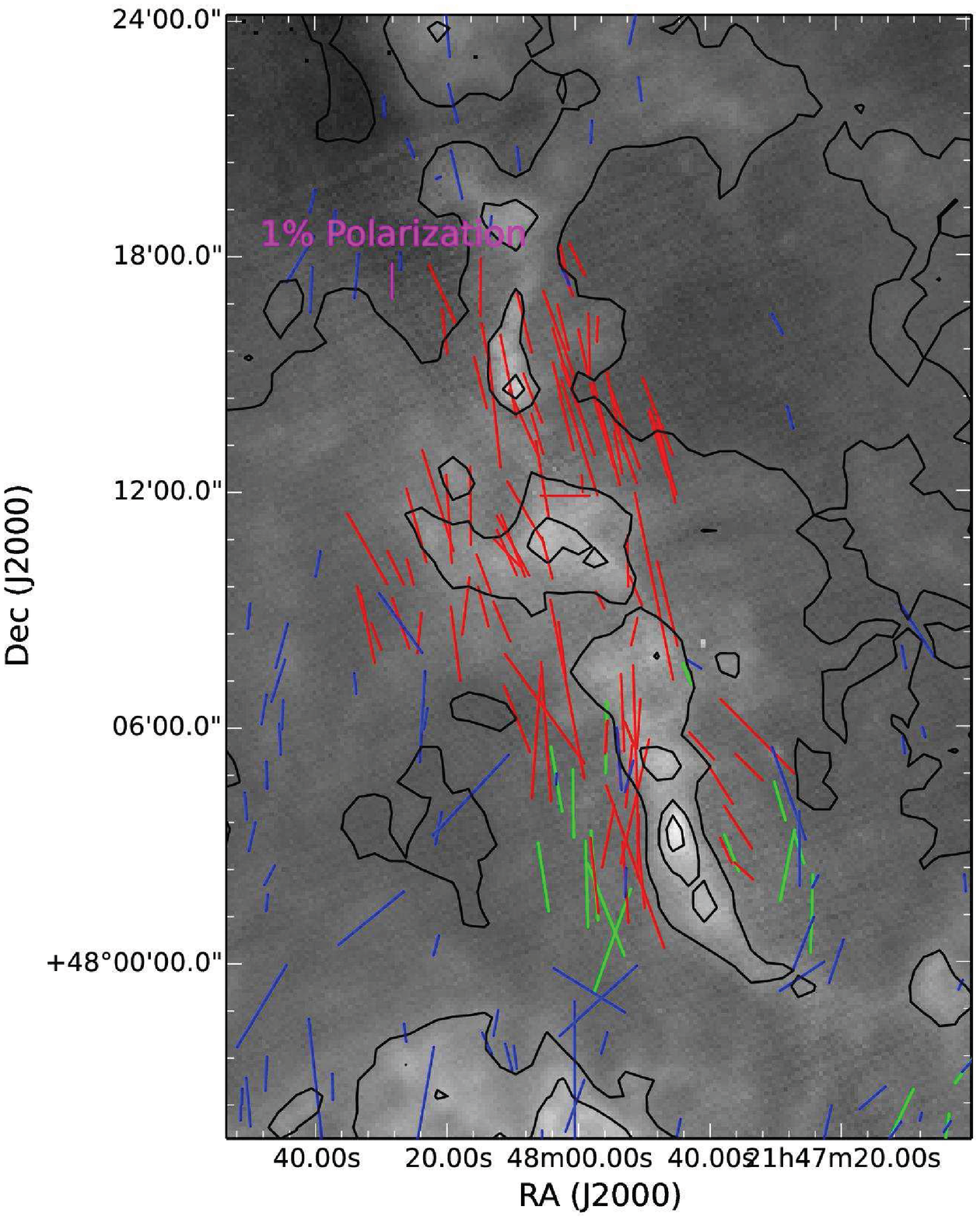}
  \caption{TRIPOL $i^\prime$-band polarization vector map (in red arrows) of IC\,5146 \citep{wan17}.  The background image is 
	  the Herschel 250~$\mu$m data \citep{arz11}.  Also shown are the polarization vectors measured by 
          AIMPOL at $R$ band (in green), and Mimir at near-infrared $H$ band (in blue).} 
  \label{fig:ic5146}
\end{figure}

We are pursuing several programs for polarimetric monitoring of Galactic star-forming regions.  
An organized polarization pattern of background stars, as the result of dichroic extinction by magnetically 
aligned dust grains, provides the magnetic field structure in a dark cloud \citep{dav51}, whereas scattered 
light reveals the radiation fields and spatial distribution of circumstellar matter of young stellar objects.  
Polarimetric observations with TRIPOL2 on the Lulin 1-m telescope were carried out for IC\,5146 on 
27 and 28 July 2012.  Seven fields were observed toward the north-west part of this filamentary cloud, 
with a total exposure time of 1.5~hour (22.5~min for each HWP angle).  Polarization measurements 
simultaneously acquired in $g^\prime$-, $r^\prime$-, and $i^\prime$-bands were corrected for both 
instrumental polarization as well as offset polarization angles by observing polarized and unpolarized 
standard stars.  Figure~\ref{fig:ic5146} shows the $i^\prime$-band polarization of the north-west 
part of IC\,5146 \citep{wan17}.  

Also plotted in Figure~\ref{fig:ic5146} are the AIMPOL (ARIES Imaging Polarimeter) 
$R$-band and Mimir $H$-band polarization data.  AIMPOL (Reutela et al. 2004), an optical polarimeter 
adapted to the 1.04~m Sampurnanand Telescope 
of the ARIES in Nainital, India, has been well calibrated over the years by observing unpolarized and 
polarized standard stars (Medhi et al. 2007, Eswaraiah et al. 2013).  
Mimir is a near-infrared imager for polarization measurements \citep{cle07} mounted 
on the 1.8~m Perkins Telescope in Arizona, operated by Lowell Observatory.

While TRIPOL2 and Mimir observations each covered a larger part of the filament than the AIMPOL data did, 
the measured polarization results by the three instruments, two working in optical and one in 
near-infrared, are consistent with each other, suggesting a global magnetic field roughly parallel to the 
long axis of the filament.  On average TRIPOL2 detected more prominent polarization than Mimir, a 
manifestation of a higher fractional polarization in visible wavelengths because the extinction difference is 
amplified.  Infrared polarimetry, on the other hand, probes denser parts of a molecular cloud.  A combination of 
optical and infrared polarimetry, together with millimeter and submillimeter interferometric observations of 
polarization, hence offers the opportunity to scrutinize the magnetic field structure from scales from 
a cloud core to the central protostar.  The detailed results of IC\,5146 can be found in \citet{wan17}.  

\begin{figure}[htb]
  \centering
  \includegraphics[width=\textwidth]{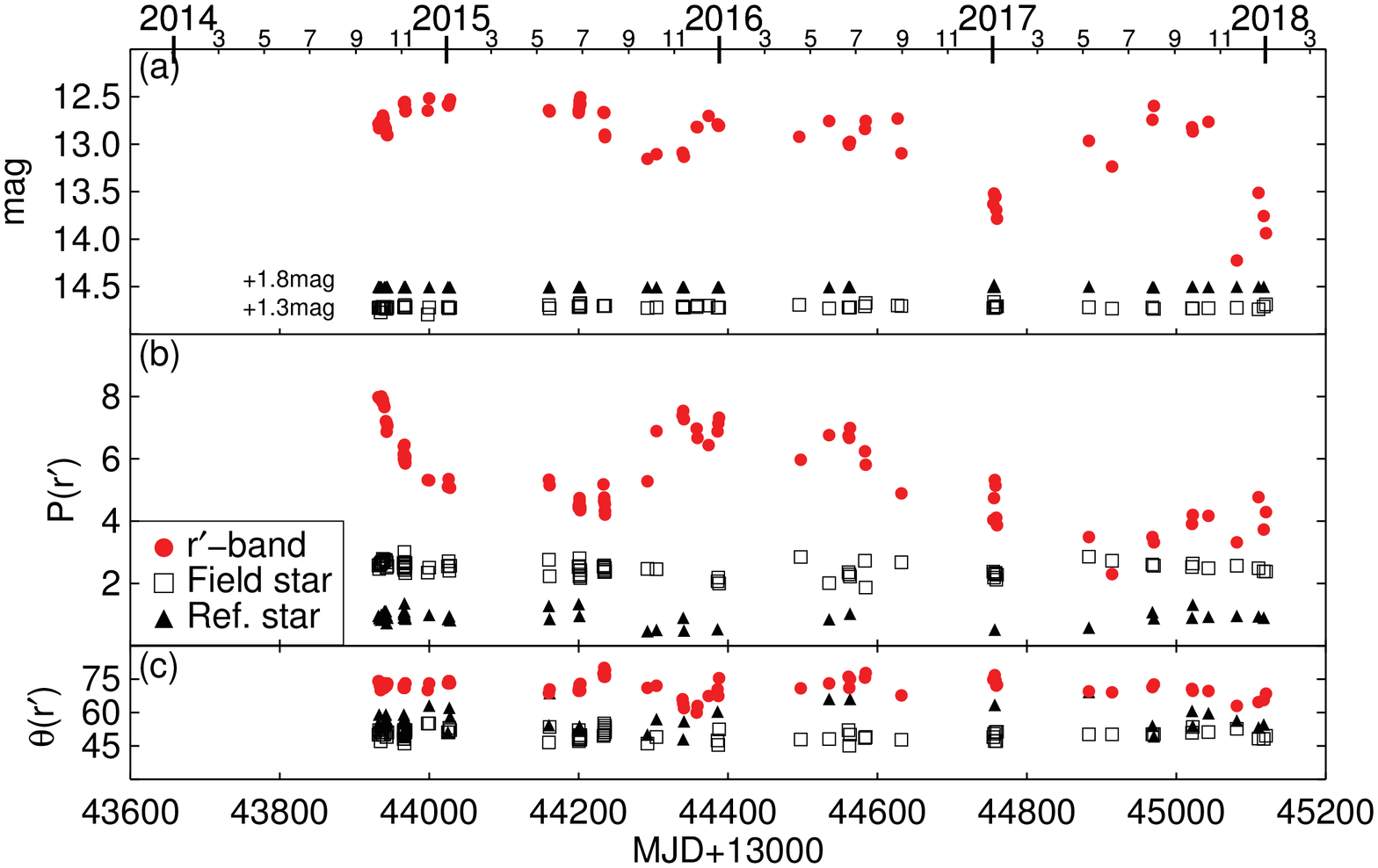}
	\caption{The $r^\prime$-band (a)~light curve, (b) polarization, and (c) polarization angle
	of the UX\,Ori-type young star GM\,Cep measured by TRIPOL from 
	late 2014 to late 2017 \citep{hua19}. While GM\,Cep shows a significant temporal change of 
	polarization in comparison with the two nearby stars .}
  \label{fig:gmcep}
\end{figure}

Another application of TRIPOL is on the point source GM\,Cep, a 4-Myr  
T~Tauri star undergoing abrupt photometric variations caused by obscuration of protoplanetary dust 
clumps \citep{che12,che14,hua19}.  The long-term photometric and polarimetric monitoring data, shown in 
Figure~\ref{fig:gmcep} show a noticeable polarization up to 8\% with temporal variability on a time scale 
of years, while the comparison star exhibits a steady level of polarization, with a standard deviation less than 1\%.  
Such polarization observations provide valuable information of the distribution and properties of the 
circumstellar dust clumps from grain growth from micron-size dust in transition to km-size planetesimals 
\citep{hua19}.

\section{Summary}

The simultaneous three-color ($g^\prime$, $r^\prime$, $i^\prime$) polarimeter, TRIPOL, is 
simple, compact and economical, suitable for a small telescope in a moderate astronomical site.  
This paper presents the design concept, and compares 
the performance to data taken on the 1-m telescope located in Taiwan.  The limiting magnitudes for 
photometry are found to be $g^{\prime}\sim19$~mag, $r^{\prime}\sim18.5$~mag, and $i^{\prime}\sim18$~mag, 
with a signal-to-noise of 10 and an integration time of 100~s.  The internal instrumental polarization is 
at the level of 0.3\% for a 100-s integration at all three bands.   
The simultaneous photometric and polarimetric capability should open up new research opportunities for 
time-domain astronomy on small or amateur telescopes. 

\begin{acknowledgements}
We are thankful to N.~Takeuchi and A.~Yamanaka 
for their laboratory experiments, and to Drs. M.~Kino, M.~Kurita, T.~Nagayama, K.~Kawabata, M.~Ishiguro, 
and H.~Takami for valuable comments throughout the development of this project.  We appreciate
Dr. Y.~Nakajima for his kind assistance for the data reduction and Drs. M.~Ishiguro and K~ Murata
for preparation of the figures.  This work is supported by Grant-in-Aid for Science 
Research from the MInistry of Education, Culture, Sports, and Technology of Japan.  
SS owes his scientific activity to Dr. H.~Fujiwara for his generous support of astronomy.
\end{acknowledgements}



\begin{table}
\begin{center}
  \caption[]{TRIPOL Measurements of Unpolarized Standard Stars} \label{tab:unpol}  
\begin{tabular}{llccc}
 \hline\noalign{\smallskip}
	Star,mag/$P_\lambda$ \citep{sch92}     & Date     & $P_{g^{\prime}} (\%) $ &  $P_{r^{\prime}} (\%) $ & $P_{i^{\prime}} (\%) $ \\
 \hline\noalign{\smallskip}
BD+$32\degr 3739$, $V=9.31$                    & 2011 Aug 14 & $0.12\pm0.11$ & $0.12\pm0.10$ & $0.17\pm0.15$  \\
~~~$P_B=0.039\pm0.021$                         & 2011 Aug 15 & $0.14\pm0.16$ & $0.32\pm0.14$ & $0.19\pm0.23$ \\ 
~~~$P_V=0.025\pm0.017$                         & 2018 Oct 25 & $0.27\pm0.27$ & $0.26\pm0.19$ & $0.32\pm0.18$ \\   
                                               & 2018 Oct 28 & $0.20\pm0.20$ & $0.16\pm0.16$ & $0.17\pm0.12$ \\   \hline
BD+$28\degr 4211$, $V=10.53$                   & 2011 Aug 15 & $0.20\pm0.19$ & $0.34\pm0.21$ & $0.25\pm0.10$   \\
~~~$P_B=0.063\pm0.023$	                       & 2011 Aug 17 & $0.08\pm0.13$ & $0.29\pm0.14$ & $0.32\pm0.28$ \\
~~~$P_V=0.054\pm0.027$                         & 2018 Oct 26 & $0.25\pm0.12$ & $0.20\pm0.20$ & $0.20\pm0.20$ \\  
~~~$P_V=0.054\pm0.027$                         & 2018 Oct 27 & $0.28\pm0.13$ & $0.17\pm0.17$ & $0.21\pm0.19$ \\  \hline
%
%
HD\,212311, $V=8.10$                           & 2018 Oct 23 & $0.15\pm0.05$ & $0.15\pm0.06$ & $0.20\pm0.05$ \\
~~~$P_B=0.028\pm0.025$                         & 2018 Oct 24 & $0.20\pm0.09$ & $0.24\pm0.06$ & $0.13\pm0.07$ \\
~~~$P_V=0.034\pm0.021$                         & 2018 Oct 25 & $0.26\pm0.12$ & $0.23\pm0.12$ & $0.28\pm0.14$ \\
                                               & 2018 Oct 26 & $0.11\pm0.11$ & $0.10\pm0.15$ & $0.32\pm0.21$ \\
                                               & 2018 Oct 27 & $0.12\pm0.12$ & $0.23\pm0.15$ & $0.16\pm0.16$ \\
                                               & 2018 Oct 28 & $0.07\pm0.11$ & $0.21\pm0.10$ & $0.12\pm0.12$ \\
%
%
%
%
%
 \hline\noalign{\smallskip}
\end{tabular}
\end{center}
\end{table}


\begin{table}
\begin{center}
	\caption[]{TRIPOL Measurements of Polarized Standard Stars    } \label{tab:pol}
\begin{tabular}{llccc}
 \hline\noalign{\smallskip}
Star/mag/$P_\lambda$, $\theta_\lambda$ \citep{sch92}  & Date     
	& $P_{g^{\prime}} (\%)$, $\theta_{g^{\prime}} (deg) $ 
	& $P_{r^{\prime}} (\%)$, $\theta_{r^{\prime}} (deg) $ 
	& $P_{i^{\prime}} (\%)$, $\theta_{i^{\prime}} (deg) $ \\
 \hline\noalign{\smallskip}
HD\,154445, $V=5.61$                            & 2015 Feb 17 & $3.8\pm0.1, 87\pm3$ & $3.4\pm0.2, 82\pm2$  & $3.7\pm0.1, 67\pm3$ \\
~~~$P_V=3.780\pm0.062$, $\theta_V=88.79\pm0.47$ & 2015 Feb 26 & $3.8\pm0.1, 92\pm3$ & $3.7\pm0.2, 90\pm2$  & $3.7\pm0.1, 92\pm3$ \\
~~~$P_{Rc}=3.683\pm0.072$, $\theta_R=88.91\pm0.56$ & 2015 Feb 27 & $3.8\pm0.1, 88\pm3$ & $4.0\pm0.2, 86\pm2$  & $3.5\pm0.1, 87\pm3$ \\
~~~$P_{Ic}=3.246\pm0.078$, $\theta_I=89.91\pm0.69$ & & & & \\
HD\,161056, $V=6.32$                            & 2015 Feb 27 & $3.9\pm0.1, 67\pm3$  & $4.1\pm0.2, 66\pm2$  & $3.7\pm0.1, 67\pm3$ \\
~~~$P_V=4.030\pm0.025$, $\theta_V=66.93\pm0.18$ & & & & \\
~~~$P_{Rc}=4.012\pm0.032$, $\theta_R=67.33\pm0.23$ & & & &\\
~~~$P_{Ic}=3.575\pm0.030$, $\theta_I=67.78\pm0.24$ & & & & \\
HD\,204827, $V=7.93$                            & 2011 Aug 11 & $5.5\pm0.2, 60\pm1$  & $5.3\pm0.2, 61\pm1$ & $4.7\pm0.2, 63\pm2$ \\
~~~$P_V=5.322\pm0.014$, $\theta_V=58.73\pm0.08$ & 2018 Oct 23 & $5.5\pm0.3, 59\pm3$  & $5.0\pm0.2, 58\pm2$ & $4.5\pm0.3, 58\pm2$ \\
~~~$P_{Rc}=4.893\pm0.029$, $\theta_R=59.10\pm0.17$ & 2018 Oct 24 & $5.8\pm0.2, 61\pm1$  & $5.3\pm0.2, 60\pm1$ & $4.5\pm0.2, 62\pm1$ \\
~~~$P_{Ic}=4.189\pm0.030$, $\theta_I=59.94\pm0.20$ & 2018 Oct 25 & $5.9\pm0.1, 59\pm1$  & $5.3\pm0.1, 60\pm1$ & $4.5\pm0.1, 60\pm1$ \\
                                                & 2018 Oct 26 & $5.9\pm0.1, 57\pm1$  & $5.2\pm0.0, 59\pm1$ & $4.2\pm0.1, 60\pm1$ \\
                                                & 2018 Oct 27 & $5.6\pm0.2, 57\pm1$  & $5.3\pm0.2, 58\pm1$ & $4.4\pm0.2, 58\pm1$ \\
                                                & 2018 Oct 28 & $5.6\pm0.1, 59\pm1$  & $5.1\pm0.1, 59\pm1$ & $4.3\pm0.1, 60\pm1$ \\ \hline
HD\,19820, $V=7.11$                              & 2018 Oct 23 & $4.5\pm0.1, 115\pm1$ & $4.4\pm0.1, 115\pm1$ & $4.2\pm0.1, 114\pm1$  \\
~~~$P_V=5.322\pm0.014$, $\theta_V=114.93\pm0.08$ & 2018 Oct 24 & $4.6\pm0.2, 114\pm2$ & $4.7\pm0.1, 111\pm1$ & $4.0\pm0.1, 115\pm2$  \\
~~~$P_{Rc}=4.893\pm0.029$, $\theta_R=114.46\pm0.17$ & 2018 Oct 25 & $4.9\pm0.2, 110\pm2$ & $4.0\pm0.2, 115\pm2$ & $4.5\pm0.1, 117\pm2$  \\
~~~$P_{Ic}=4.189\pm0.030$, $\theta_I=114.48\pm0.20$ & 2018 Oct 26 & $4.2\pm0.1, 111\pm2$ & $3.8\pm0.1, 113\pm1$ & $3.5\pm0.1, 115\pm2$  \\
~~~(Variable, this work)                                       & 2018 Oct 27 & $4.4\pm0.1, 111\pm1$ & $4.3\pm0.1, 113\pm1$ & $3.6\pm0.1, 115\pm1$  \\
                                                 & 2018 Oct 28 & $4.6\pm0.2, 114\pm1$ & $4.6\pm0.2, 113\pm1$ & $4.0\pm0.1, 113\pm1$  \\ \hline
 \hline\noalign{\smallskip}
\end{tabular}
\end{center}
\end{table}


\label{lastpage}

\end{document}